# Performance Evaluation of Hierarchical Publish-Subscribe Monitoring Architecture for Service-Oriented Applications

Ivan Zuzak
University of Zagreb
Faculty of Electrical Engineering and Computing
Zagreb, Croatia
izuzak@gmail.com

Ivan Benc
Croatian Telecom
Zagreb, Croatia
ivanbenc@gmail.com

*Abstract* - **Contemporary high-performance service-oriented applications demand a performance efficient run-time monitoring. In this paper, we analyze a hierarchical publish-subscribe architecture for monitoring service-oriented applications. The analyzed architecture is based on a tree topology and publish-subscribe communication model for aggregation of distributed monitoring data. In order to satisfy interoperability and platform independence of service-orientation, monitoring reports are represented as XML documents. Since XML formatting introduces a significant processing and network load, we analyze the performance of monitoring architecture with respect to the number of monitored nodes, the load of system machines, and the overall latency of the monitoring system.**

## I. INTRODUCTION

*Service-oriented computing* (SOC) [1] is established as a major paradigm for creating distributed applications. In service-oriented computing, applications are created using loosely coupled, dynamically bounded, platform independent program modules called *services*. As the number, size and complexity of service-oriented applications increases [2], supervision and management of these systems becomes performance demanding. Moreover, the platform independence, interoperability, and loose coupling of SOC result in difficult run-time collection and aggregation of large amounts of distributed data [3]. Most notably, interoperable service execution in heterogeneous environments requires platform-independent data exchange. However, XML, which is commonly used for data serialization, is inefficient with respect to message size, message transfer rates and imposed processing load [4]. Therefore, performance tradeoffs need to be made during the design of systems for monitoring service-oriented applications.

In this paper, we analyze the performance implications of hierarchical monitoring in service-oriented applications. We focus our analysis on limitations and necessary performance tradeoffs of monitoring applications based on WebServices protocol stack [5]. The analyzed monitoring architecture is based on hierarchical aggregation of XML documents containing reports collected from sensor services. The hierarchical aggregation process is achieved by utilizing publish-subscribe mechanisms organized in a tree topology. For modeling and implementation of a monitoring system based on the presented architecture, we use the Service-Oriented Programming Model (SOPM) [6]. SOPM enables the design, development, and execution of service-oriented applications in which services mutually cooperate and compete without central control.

In order to evaluate the architecture's performance we utilize a hybrid approach based on an analytic model upgraded with results of measurements on implemented systems. For example, we estimate the impact of the number of sensor services, their monitoring reporting frequencies, and monitoring report sizes on the load of system machines, and overall latency of the monitoring system based on measurement results. Since hierarchical topologies are commonly used in data aggregation systems [7] [13], our evaluation is applicable to a large number of existing monitoring systems.

The remainder of the paper is organized as follows. An overview of related research in the field of large-scale distributed monitoring is given in section II. In Section II, we also describe the Service-Oriented Programming Model. The analyzed monitoring architecture is presented in section III, while section IV presents the results of analytical and empirical performance evaluation. Section V concludes the paper with an analysis of the evaluation and gives recommendations for usage of hierarchical monitoring systems.

## II. BACKGROUND

### A. Large-scale distributed monitoring systems

One of more recent research efforts in distributed system monitoring is the *Grid Monitoring Architecture* (GMA) [7], a generic monitoring model which is not constrained by protocol usage or by the underlying data model. GMA is often used in development of large-scale service-oriented monitoring systems. In GMA, *Consumer* components query *Registry* components to find out what type of monitoring information is available in the system, locate *Producer* components that provide this information, and contact *Producers* to obtain the relevant information.

Based on the GMA architecture, a four level taxonomy of monitoring systems was developed [8]. *Level 0* systems are monolithic in the way that producers are implemented within the same system as consumers, while in *Level 1* systems producers are implemented separately. *Level 2* systems introduce republisher components which implement both producer and consumer interfaces. Although these components may be distributed and replicated, republishers in *Level 2* systems are constrained with respect to their interconnection and network organization. Finally, *Level 3* systems introduce configurable republishers, allowing their organization in arbitrarily hierarchical structures. In such hierarchies, republishers collect events from lower level producers and aggregate them into higher level events. The architecture of monitoring systems analyzed in this paper allows arbitrarily structured data aggregation hierarchies and may be considered a *Level 3* GMA system.

The GMA architecture has been implemented in many existing large-scale systems. Although the performance of certain implementations has been evaluated, most of these evaluations were not comprehensive. Most evaluations were performed only experimentally and only for several specific system configurations. Furthermore, the experiments were performed by varying only a single system parameter, such as the number of producers, and measuring the performance only by observing a single system property, such as the overhead introduced on CPU or network.

In [9] a Java-based design pattern of Grid Monitoring System is presented and implemented. The network traffic overhead introduced by the monitoring system is studied for a single system configuration and varying message size. Furthermore, for the implemented system the authors experimentally conclude the number of Producers required to saturate a Consumer. The evaluation of the system is continued in [16] where the authors also provide measurements of the systems' response time as a function of the message size.

In [10] GridEye: A Service-oriented Grid Monitoring System is presented and tested at the China National Grid. Specific monitoring traces are analyzed in order to determine the CPU, memory and disk usage overhead on monitoring nodes.

In [17] the design and implementation of a GMA based Grid Monitoring Service is presented. Although the paper does not provide any evaluation results, the authors derive an architecture-level conclusion that the collecting frequency is the key factor to the precision of the monitor information and the load which is introduced to the system by sensors and that a contradiction between load and precision exists.

Ganglia [18] is a popular distributed monitoring system for high performance computing systems such as clusters and Grids. Although the performance and scalability evaluation of the system is extensive, it is still performed through experimental measurements. For performance overhead, the CPU, memory footprints, and network bandwidth was measured. For scalability, the overhead on individual nodes was measured and quantified how overhead scales with the size of the system, both in terms of number of nodes within a cluster and the number

The evaluation presented in this paper is an architecture-level performance analysis, i.e. the architecture is evaluated for various system configurations and with respect to both different system parameters and performance criteria. Furthermore, the evaluation is analytical as well as experimental which enables insight into system performance a-priori to system execution.

*B. Service Oriented Programming Model*

Service-Oriented Programming Model (SOPM) [6] is a methodology for the design, development, and execution of service-oriented applications. SOPM applications consist of *Application services*, *Coopetition services* and *Distributed programs*. Application services implement coarse fragments of application's computational logic while Coopetition services are pre-built services of the SOPM environment for coordination and synchronization of Application services. Distributed programs, specified in a process description language CL [12], use Coopetition services to bind Application services into a distributed application.

The monitoring system presented in this paper utilizes the EventChannel Coopetition service [11]. EventChannel is an extension of the classic publish-subscribe mechanism into a publish-subscribe-interpret mechanism that supports development of event-driven document-oriented distributed systems. In order to use the EventChannel, *Publisher Distributed programs* publish events on the *EventChannel* while *Subscriber Distributed programs* subscribe to sets of events. The process of matching published events to active subscriptions is implemented by external, application specific *Interpreter Services*. The *EventChannel* sends a notification to Subscriber Programs whenever published events match a subscription. Additionally, before forwarding events to subscribers, the Interpreter service may modify the event set. We use this feature of the Interpreter service to implement the aggregation of multiple monitoring report documents into a single document.

### III. HIERARCHICAL PUBLISH-SUBSCRIBE MONITORING ARCHITECTURE

The monitoring system aggregates monitoring reports from a set of distributed sensor services. The aggregation process creates a single global report combined from reports of all system sensors [14]. In order to support scalable execution of this process, the architecture of monitoring system is based on multi-level hierarchical aggregation where multiple reports from a lower level are combined into a single report on the higher level.

Figure 1 presents the process of hierarchical aggregation of monitoring reports. On the lowest level, **service reports** contain data from a single application service, while **node reports** aggregate reports from application services executing on the same machine. Node reports from different machines are aggregated into **intermediate reports** through several levels of the hierarchy, and finally they are combined into a single **distributed system report**. Since reports on a higher level contain all information from lower level, their size grows as the level of hierarchy increases. Furthermore, in order to comply with platform neutrality of service-oriented systems, reports are implemented as XML documents.

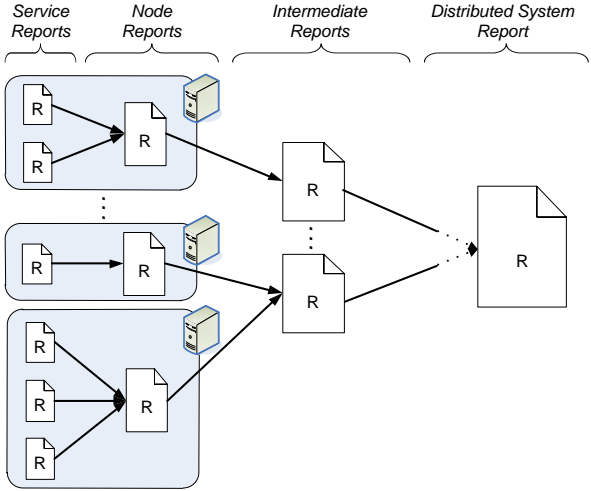

Figure 1. Aggregation of monitoring reports

In order to support loose coupling of system services, the monitoring architecture is based on SOPM *EventChannel* publish-subscribe-interpret mechanisms for collecting reports. As presented in the Figure 2, the system is organized into multiple levels: the service level, intermediate aggregation levels and the distributed system level. On the **service level**, each system machine runs a *Sensor service* that periodically collects service reports from *Application services*, creates a node report and publishes it to the first aggregation level.

Each **aggregation level** contains several **Monitoring EventChannels**, each with an associated **Aggregator service** that operates as the Interpreter service and **Forwarder service** which operates both as the Subscriber program of the current level and Publisher program for the next level. Monitoring EventChannels receive reports from the lower level and forward received reports to Aggregator services that periodically aggregate them into a single intermediate report. The Monitoring EventChannel and associated Aggregator and Forwarder services are executed on the same machine. The number of Monitoring EventChannels decreases as the level of aggregation increases.

Intermediate reports are forwarded to the Forwarder service that publishes it to a Monitoring EventChannel on the next level. The same process occurs on the **distributed system level** that contains a single Monitoring EventChannel. The system-level Aggregator service periodically aggregates received intermediate reports into a single distributed system report that is forwarded to the distributed system level Forwarder service.

## IV. PERFORMANCE ANALYSIS OF THE HIERARCHICAL PUBLISH-SUBSCRIBE MONITORING ARCHITECTURE

The major limiting factor of the performance of the architecture is the use of XML formatting which introduces high overhead on system processing and network transfer load. In order to achieve acceptable system operation, a tradeoff analysis is required to reveal the influence of variable aspects on the system performance. For example, a possible tradeoff could be between distributed system report freshness and reporting period.

In order to investigate how architecture attributes, such as number of sensors, reporting frequencies and report sizes, affect the performance, we create an analytic model and perform empirical measurements on a prototype implementation. The combination of the analytic model and measurements determine limitations for various architecture configurations. The following metrics are of special interest: (a) the maximum number of machines that can be monitored by the proposed monitoring architecture, (b) the propagation time of a report to the distributed system level, and (c) the load on machines imposed by monitoring activities.

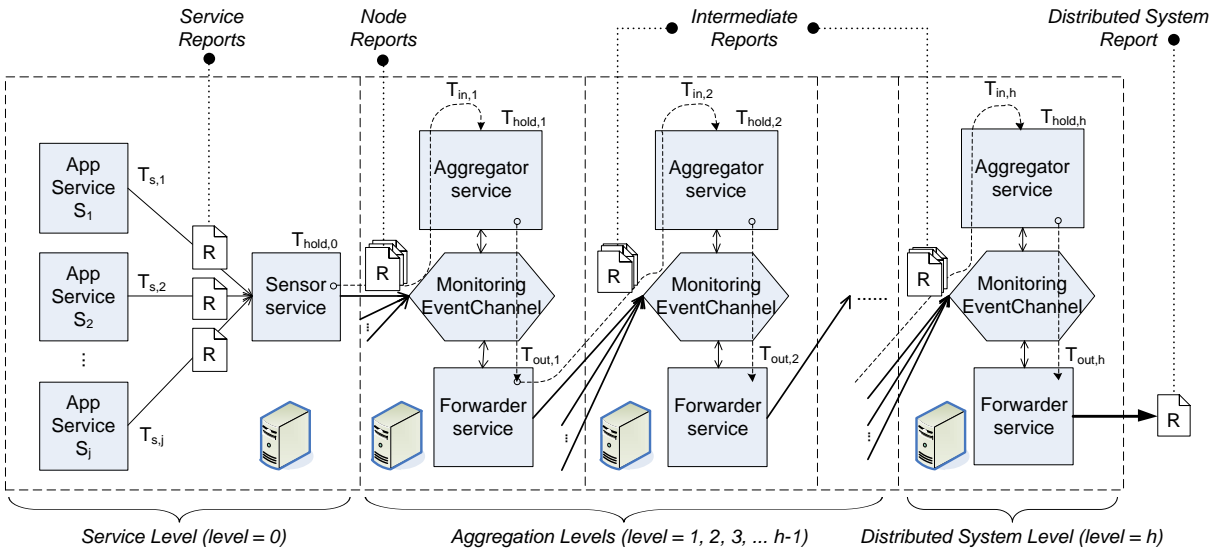

Figure 2. Hierarchical publish-subscribe monitoring architecture

## A. Analytic model

In order to determine the time needed to propagate a service report to the distributed system level Aggregator service, we propose an analytic model of the architecture. To simplify the analysis, the analytic model assumes that every Monitoring EventChannel at a specific level receives reports from the same number of lower level services, i.e. the tree representing the topology of the system is balanced. This assumption is justified since balanced trees are common in tree-based aggregation topologies due to their even distribution of load on system nodes [13].

The purpose of the analytic model is to estimate the propagation times ($T_{prop}$) and staleness times ($T_{stale}$) of monitoring reports, as defined by the parameters of the architecture in Table 1 and shown in Fig 2.

TABLE I.    PARAMETERS OF THE HIERARCHICAL PUBLISH-SUBSCRIBE MONITORING ARCHITECTURE

| Term | Meaning |
| --- | --- |
| h | The depth of hierarchy represented by the number of aggregation levels plus one for the service level. |
| $n_i$ | The number of services sending monitoring reports to the Monitoring EventChannel at level **i**. For i=0, $n_i$ represents the number of different Application services sending reports to a Sensor service. |
| $T_{s,j}$ | The reporting period in which Application service $S_j$ emits a service report regarding its performance. |
| $T_{in,i}$ | The *input time* on level **i**. This is the time it takes a report to propagate from the Forwarder service on level **i-1** to the Aggregator service of level **i**. |
| $T_{hold,i}$ | The *holding time* of the Aggregator service on level **i**. This is the period during which the Aggregator service accumulates new reports before sending them to the Forwarder service. |
| $T_{out,i}$ | The *output time* on level **i**. This is the time it takes a report to propagate from the Aggregator service to the Forwarder service. |
| $T_{prop,j}$ | The *propagation time* for Application service $S_j$. This is the time it takes a service report from $S_j$ to be formed and to propagate to the distributed system level Aggregator service. The value of $T_{prop,j}$ is calculated through a series of $T_{prop,j,i}$ values which represent the times it takes an application condition to reach the **i**th level Aggregator service of the hierarchy. $T_{prop,j,0}$ represents the time required for a condition in an Application service to reach the Sensor service. |
| $T_{stale,j}$ | The *staleness time* of service reports defined as the upper limit of the oldness of reports from Application service $S_j$ in the distributed system level Aggregator service. The value of $T_{stale,j}$ is calculated through a series of $T_{stale,j,i}$ which represent the staleness of reports at the $i^{th}$ level of the hierarchy, analogous to the computation of $T_{stale,j,i}$. |

The analytic model is represented with the following equations. The propagation time of a report in Application service $S_j$ to the Sensor service is represented with expression (1):

$$T_{prop,j,0} = T_{hold,0}, j \in [0, n_0] \quad (1)$$

The propagation time $T_{prop,j,0}$ is defined by the time it takes a Sensor service to pick up the service report, which is in the worst case $T_{hold,0}$. The propagation time for the first-level Monitoring EventChannel is represented with expression (2):

$$T_{prop,j,1} = T_{prop,j,0} + T_{in,1}, j \in [0, n_0] \quad (2)$$

The propagation time $T_{prop,j,1}$ is defined by the propagation time for the Sensor service ($T_{prop,j,0}$) incremented by the time it takes a report to be received by the Aggregator service ($T_{in,1}$). Expression (3) represents the way to calculate the time it takes a report of an Application service $S_j$ to reach an Aggregator service of the $i^{th}$ level Monitoring EventChannel:

$$T_{prop,j,i} = T_{prop,j,i-1} + T_{hold,i-1} + T_{out,i-1} + T_{in,i}, \\ j \in [0, n_0], i \in [2, h] \quad (3)$$

As formulated, the maximum time it takes to reach a particular Aggregator service is equal to the maximum time it takes to reach the lower level Aggregator service, plus the maximum waiting time in the lower level Aggregator service ($T_{hold,i-1}$), plus the time it takes a report to be received at the lower level Forwarder service ($T_{out,i-1}$), plus the time it takes to forward the report from the lower level Forwarder service to the next level Aggregator service ($T_{in,i}$).

Expression (4) is obtained by recursively solving expression (3) and using expressions (1) and (2) for $T_{prop,j,0}$ and $T_{prop,j,1}$, respectively:

$$T_{prop,j,i} = \sum_{k=0}^{i-1} T_{hold,k} + \sum_{k=1}^{i-1} T_{out,k} + \sum_{k=1}^{i} T_{in,k}, j \in [0, n_0], i \in [2, h] \quad (4)$$

Expression (5) represent the way to calculate the staleness time of a report from Application service $S_j$ on the $i^{th}$ level Monitoring EventChannel:

$$T_{fresh,j,i} = T_{reach,j,i} + T_{s,j}, j \in [0, n_0], i \in [2, h] \quad (5)$$

Since in the worst case a report of service $S_j$ is $T_{s,j}$ old, the staleness time is calculated from the time it takes a condition to be represented in a custom service report ($T_{s,j}$), plus the time it takes the report to propagate to the $i^{th}$ level Monitoring EventChannel ($T_{prop,j,i}$).

Parameters $T_{hold,i}$ and $T_{s,j}$ are constant for a given monitoring system configuration. However, values $T_{out,i}$ and $T_{in,i}$ depend on the system and network load. Therefore, we perform measurements to determine values of $T_{out,i}$ and $T_{in,i}$ for various system conditions.

## B. Empirical results

In this section, we present results obtained through experimental measurements and simulation of the monitoring system for each level in the system architecture. We use these results for estimating the $T_{out,i}$ and $T_{in,i}$ parameters in the overall system analysis.

In general, monitoring reports vary in size depending on the number and type of Application services running on each machine. However, for empirical measurements we have used node reports of fixed size of 0.5 kB.

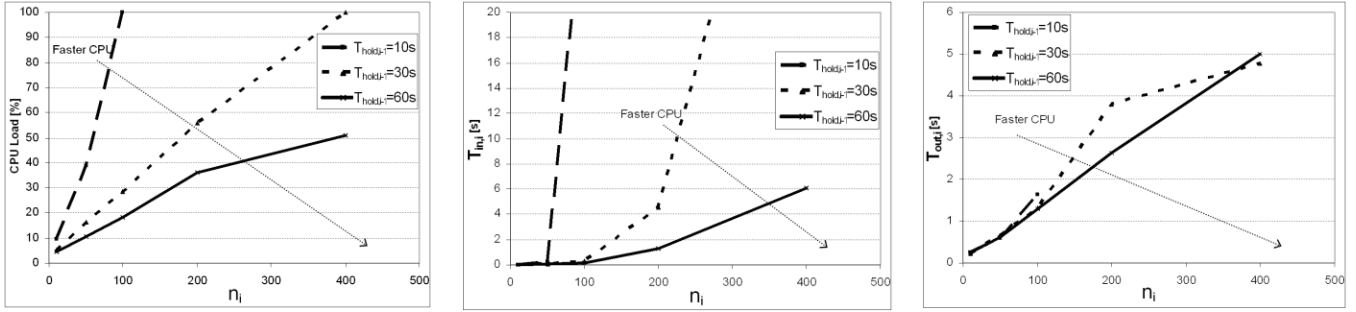

Figure 3. CPU load, $T_{in,i}$ and $T_{out,i}$ of the EventChannel machine at $i^{th}$ level as a function of the number of $(i-1)^{th}$ level services that are sending reports and their reporting rates. $T_{hold,i}$ = 30s.

*1) Sensor service*

When analyzing the Sensor service two parameters are considered: the reporting period in which reports are published to the Monitoring EventChannel ($T_{hold,0}$), and the reporting period of Application services on the local node ($T_{s,j}$). Regarding the number of machines that can be simultaneously monitored, the Sensor service does not impose any restrictions, since it is started and executed independently on each system machine. On the other hand, the execution of Sensor services affects the CPU load of local machines. Therefore, the reporting period $T_{hold,0}$ should not be too small, since high reporting frequencies can have a significant effect on the CPU load.

*2) Monitoring EventChannels*

When analyzing Monitoring EventChannels on $i^{th}$ level several parameters need to be considered: the number of lower level services sending reports ($n_i$), holding times of both lower level Aggregator service ($T_{hold,i-1}$) and current level Aggregator service ($T_{hold,i}$), and the size of reports that are sent from lower level services.

Fig. 3 shows the CPU load, input time $T_{in,i}$, and output time $T_{out,i}$ of one of the Monitoring EventChannel machine at $i^{th}$ level as a function of the number of $(i-1)^{th}$ level services ($n_i$) and their holding times ($T_{hold,i-1}$) in seconds. Increase in the number of services sending reports increases the CPU load of the Monitoring EventChannel machine. Furthermore, as the frequency with which services send reports increases ($1/T_{hold,i-1}$), the CPU load of the Monitoring EventChannel also increases. Moreover, results indicate that the input time $T_{in,i}$ increases gradually until the CPU goes to saturation. After that the input time increases infinitively. Lastly, the output time $T_{out,i}$ increases approximately linearly with the number of lower level services that are sending reports.

Fig. 4 shows the correlation of the report size (r) that is sent from services on level i-1, and the number of services on level i-1 ($n_i$) on the CPU load, input time and the output time of the $i^{th}$ level Monitoring EventChannel. The report size is expressed as the number of node reports which it contains, e.g. for r=30 the report contains 30 node reports. As can be seen, the CPU load is significantly affected by the size of the reports. Consequentially, the input time $T_{in,i}$ of the Monitoring EventChannel grows significantly as the CPU load approaches 100%. However, unlike with the previous scenario, the output time $T_{out,i}$ is significantly affected by both the report size and the number of lower level services that are sending reports. The observed result is a consequence of the output report size growing linearly with the number of lower level services.

*3) Distributed system level EventChannel*

Since the distributed system level may be observed as an intermediate level with only one Monitoring EventChannel, measurements conducted for the system level Monitoring EventChannel have similar results as those for the aggregation levels. The CPU load of the distributed system level EventChannel machine increases with the increase in the number of services that are sending reports to the distributed system level Aggregator service, with higher report sending frequencies the load of the system level machine grows faster. This indicates that a significant parameter in the system is the input frequency at the system

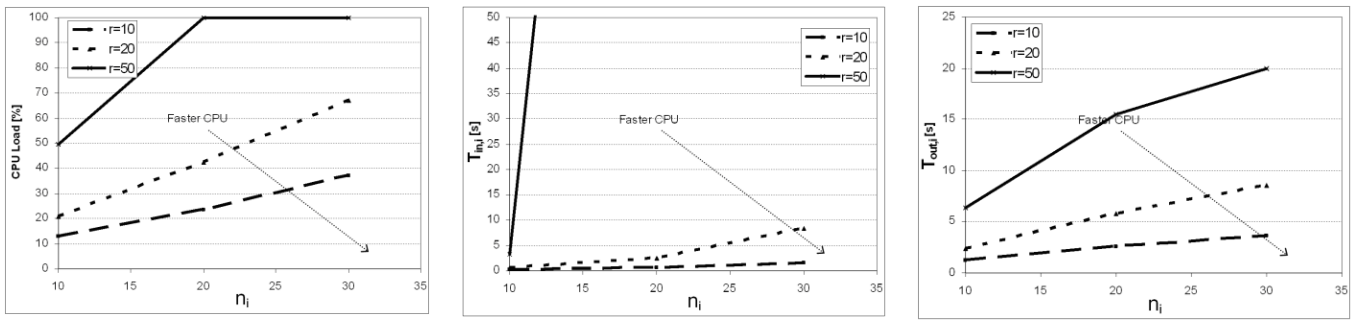

Figure 4. CPU load, $T_{in,i}$ and $T_{out,i}$ of the EventChannel machine at $i^{th}$ level as a function of the number of $(i-1)^{th}$ level services that are sending reports and their report size. $T_{hold,i}$ = 30s and $T_{hold,i-1}$ = 30s.

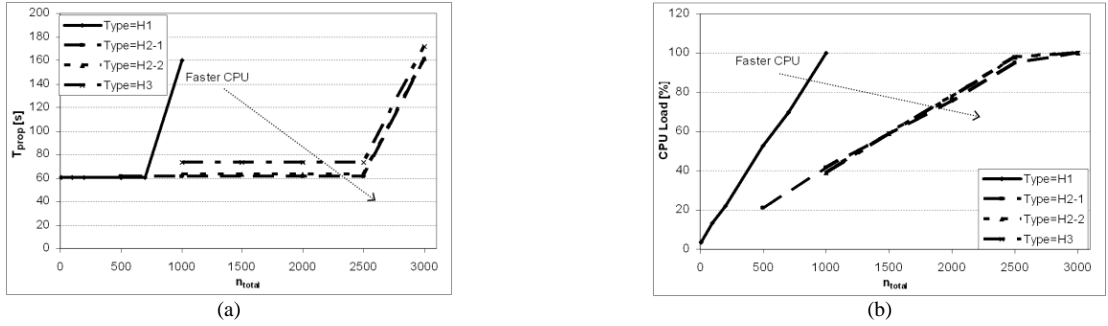

Figure 5. The limits of the system that can be monitored with different hierarchies: (a) propagation time, (b) distributed system level machine CPU load

level EventChannel which is proportional to both the number of lower level services sending reports and the frequency with which they send reports. When observing the input time of the distributed system level EventChannel, we find that it is negligible as long as the CPU load of the machine executing the system level EventChannel is not close to 100%. After the CPU goes into saturation, the input time grows significantly.

Furthermore, increase of the CPU load depends on the size of reports sent from lower level services. As the size of the reports increases, the CPU load grows faster. Additionally, the input time of the distributed system level Aggregator service is approximately constant until the CPU load approaches 100%, after which it grows rapidly. Unlike with previous results, while the CPU load is lower then 100% the input times for different report sizes are diverse. This is a result of larger reports requiring more time for parsing no matter the CPU load. Furthermore, with larger reports the CPU load goes to saturation faster, and thus the input time starts to grow significantly faster for configurations with larger reports.

*C. Overall analysis*

The purpose of the overall analysis is to determine the limits of applicability of hierarchical monitoring with the maximum number of machines that can be monitored as the most important limit. In order to explore these limits, we combine the analytic model and empirical results presented in previous sections to estimate the performance attributes of four different hierarchies. Table 2 presents the experimental hierarchies for which we performed estimations.

TABLE II. EXPERIMENTAL HIERARCHIES

| Name | Hierarchy structure and parameters | |
|---|---|---|
| Single-level hierarchy | Sensor services on a single Monitoring EventChannel (h=1). Reporting periods $T_{hold,0}$ of Sensor services is 60s. | |
| Two-level hierarchy$_1$ | Sensor services, multiple intermediary EventChannels and a system level EventChannel (h=2). $T_{hold,0}$ and $T_{hold,1}$ are 30s. The number of first level EventChannels is varied. | 50 Sensor services connected to first level EventChannel. |
| Two-level hierarchy$_2$ | | 100 Sensor services connected to first level EventChannel. |
| Three-level hierarchy | Three level hierarchy (h=3) with groups of 10 Sensor services ($n_0$=10) sending reports to first level every 10s, groups of 10 first level EventChannels ($n_1$=10) forwarding reports to second level every 30s, and second-level EventChannels forward the reports to system level every 30s. The number of second level EventChannels is varied. | |

Fig. 5 presents the comparison of the hierarchies with respect to the total number of machines in the system ($n_{total}$). The maximum propagation time ($T_{prop}$) of a report to the distributed system level Aggregator service is presented on Fig. 5 a). For the single-level hierarchy propagation time starts to grow significantly even for a low number of machines. However, the results also indicate that there is no significant benefit of creating a three-level hierarchy over two-level hierarchies. Also, there is no significant difference in performance between the two two-level hierarchies. Fig. 5 b) indicates that the CPU load of the distributed system level machine grows linearly with the growth of the number of services sending reports until it reaches 100%. Furthermore, the total propagation time grows rapidly as the distributed system level machine approaches 100% CPU load.

These results demonstrate that the hierarchy of Monitoring EventChannels is limited due to the fact that the distributed system level Aggregator service must parse all the reports coming from all machines being monitored. Therefore, simple grouping of reports into larger report by using hierarchy has limited applicability. For instance, for the given scenarios the only meaningful hierarchies are two-level hierarchies that can extend the size of the monitored system approximately five times.

V. CONCLUSION

We are witnessing an increase in the scale and complexity of service-oriented distributed applications. In order to control and supervise these applications effectively, efficient and scalable monitoring systems are needed. These scalability requirements are gaining importance even more as service-oriented computing technologies are increasingly used as an execution platform on hardware-constrained devices [15].

In this paper we analyze a publish-subscribe architecture for hierarchical aggregation of distributed monitoring reports of service-oriented applications. The analyzed architecture is based on a tree topology which is often used for distributed data aggregation. In the analyzed architecture, the leaves of the topology are sensor services that collect XML reports from individual system machines and publish them periodically to their parents in the topology. The intermediary nodes of the topology are machines that aggregate the reports from lower levels and publish them to their parents as a single report representing the state of a larger part of the system. Finally, the root machine of the topology creates a global, distributed system report about the state of the monitored system.

In order to evaluate the limits of the architectures' efficient performance we present an analytic model of the architecture and conduct empirical measurements on a prototype implementation. Together, these form the basis of an architecture-level performance analysis which we suggest for future evaluations of large-scale monitoring architectures. In our analysis we focus on exploring architecture-level limits introduced with the use of the XML language for platform independent monitoring report serialization.

First, analysis results shows that the limiting factors to the scalability regarding the size of the monitored system are the frequency of publishing reports, the size of the reports and the CPU power of the root machine. We show that by varying the frequency of publishing reports, a tradeoff may be achieved between the number of the machines that are monitored and the staleness of the global, distributed system reports. Lower reporting frequencies enable a higher number of machines to be monitored, but at the price of higher staleness of reports, and vice versa. Similarly, by changing the size of reports we can achieve a tradeoff between the number of machines that may be monitored and the quality of the reports.

Furthermore, the size of the monitored system is constrained by the CPU of the root machine since it processes reports that contain all the data from the system regardless of the exact topology. However, in our analysis we show that benefits can be achieved by reorganizing the tree topology, most notably varying between a deeper tree and a wider tree. We show that the largest benefit comes from introducing the first level of hierarchy, while introducing additional levels does not significantly increase the possible size of the monitored system.